\documentclass[conference]{IEEEtran}

\makeatletter
\newcommand{\newlineauthors}{%
  \end{@IEEEauthorhalign}\hfill\mbox{}\par
  \mbox{}\hfill\begin{@IEEEauthorhalign}
}
\makeatother

\usepackage{cite}
\usepackage[cmex10]{amsmath}
\usepackage{array}
\usepackage{mdwmath}
\usepackage{mdwtab}
\usepackage{eqparbox}

\usepackage{graphicx}
\graphicspath{{./}{figures/}}

\usepackage{caption}
\usepackage{subcaption}

\usepackage{fixltx2e}
\usepackage{stfloats}
\usepackage{url}
\usepackage{footnote}
\makesavenoteenv{tabular}
\makesavenoteenv{table}

\usepackage{booktabs}

\usepackage{xcolor}

\usepackage{makecell}

\usepackage{amsthm}
\newtheorem{theorem}{Theorem}

\usepackage{xspace}

\newcommand{\bitcoin}{Bitcoin\xspace}

\newcommand{\prism}{Prism\xspace}
\newcommand{\ethereum}{Ethereum\xspace}
\newcommand{\libra}{Libra\xspace}
\newcommand{\evm}{EVM\xspace}
\newcommand{\movevm}{MoveVM\xspace}
\newcommand{\evmprism}{{\it EVM Prism}\xspace}
\newcommand{\movevmprism}{{\it MoveVM Prism}\xspace}
\newcommand{\prismconsensusonly}{{\it Prism Consensus Only}\xspace}
\newcommand{\evmexecutoronly}{{\it EVM Executor Only}\xspace}
\newcommand{\movevmexecutoronly}{{\it MoveVM Executor Only}\xspace}
\newcommand{\donothing}{{\it Do Nothing}\xspace}
\newcommand{\nativepayment}{{\it Native Payment}\xspace}
\newcommand{\erc}{{\it ERC20}\xspace}
\newcommand{\cryptokitties}{{\it CryptoKitties}\xspace}
\newcommand{\cpuheavy}{{\it CPU Heavy}\xspace}
\newcommand{\ioheavy}{{\it IO Heavy}\xspace}
\newcommand{\vmexecutoronly}{{\it VM Executor Only}\xspace}
\newcommand{\prismconsensus}{Prism Consensus\xspace}

\usepackage{cleveref}
\crefformat{section}{\S#2#1#3}
\crefformat{subsection}{\S#2#1#3}
\crefformat{subsubsection}{\S#2#1#3}
\crefformat{footnote}{#2\footnotemark[#1]#3}

\begin{document}
%
\title{\prism Removes Consensus Bottleneck\\ for Smart Contracts}

\author{\IEEEauthorblockN{Gerui Wang}
\IEEEauthorblockA{Department of Computer Science\\University of Illinois Urbana-Champaign\\
Urbana, USA\\
Email: geruiw2@illinois.edu}
\and
\IEEEauthorblockN{Shuo Wang}
\IEEEauthorblockA{Department of Electrical Engineering\\Stanford University\\
Stanford, USA\\
Email: shuow94@stanford.edu}
\and
\IEEEauthorblockN{Vivek Bagaria}
\IEEEauthorblockA{Department of Electrical Engineering\\Stanford University\\
Stanford, USA\\
Email: vbagaria@stanford.edu}
\newlineauthors
\IEEEauthorblockN{David Tse}
\IEEEauthorblockA{Department of Electrical Engineering\\Stanford University\\
Stanford, USA\\
Email: dntse@stanford.edu}
\and
\IEEEauthorblockN{Pramod Viswanath}
\IEEEauthorblockA{Department of Electrical and Computer Engineering\\University of Illinois Urbana-Champaign\\
Urbana, USA\\
Email: pramodv@illinois.edu}
}



\maketitle

\begin{abstract}
The performance of existing permissionless smart contract platforms such as \ethereum is limited by the consensus layer. \prism \cite{prism-theory} is a new proof-of-work consensus protocol that provably achieves throughput and latency up to physical limits while retaining the strong security guarantees of the longest chain protocol. This paper reports experimental results from implementations of two smart contract virtual machines, \evm and \movevm, on top of \prism and demonstrates that the consensus bottleneck has been removed. Code can be found at \url{https://github.com/wgr523/prism-smart-contracts}.
\end{abstract}


%

\section{Introduction}

Existing permissionless smart contract platforms such as \ethereum is based on the longest chain consensus protocol, the original blockchain protocol invented by Nakamoto~\cite{bitcoin}.  While maintaining high security against adversarial attacks, it is well known that the longest chain protocol suffers from poor throughput and latency performance. 
Hence, the performance of these platforms is limited by the consensus layer.

This limitation has led to practical  congestion in the network; a noteworthy instance occurred when  CryptoKitties made its debut on \ethereum, a spike of transactions rushed into the system, far exceeding \ethereum's supported throughput. The pending transaction queue was growing quickly, and users had to increase transaction fees to incentivize miners to add their transactions to the chain. Decentralized Finance applications have been rapidly growing over the last few years and as it gets more popular in the near future, the demand will continue to grow, making the performance scaling of smart contract platforms an urgency. 

Several promising efforts to scale the performance have been proposed. Almost every major live smart contract platform such as \ethereum, Algorand, and Tron are optimizing their existing smart contract engines to increase the throughput. A few others like Libra (led by Facebook) and Hyperledger Fabric (led by IBM) have taken the route of permissioned blockchains to obtain higher throughput. On the other hand, \ethereum foundation has taken a sharding approach to support higher throughput. Optimistic Rollup~\cite{optimisticrollup}, ZK-Rollup~\cite{zkrollup}, and Arbitrum~\cite{kalodner2018arbitrum} are other off-chain scaling solutions built on top of an existing smart contract platform such as \ethereum. 
In these off-chain solutions, not every validator node needs to keep track of the execution of the off-chain contracts, which leads to an improved overall efficacy but at the expense of security. 




\begin{figure}[!htb]
    \centering
    \begin{subfigure}[b]{\columnwidth}
    \centering
    \includegraphics[width=\columnwidth]{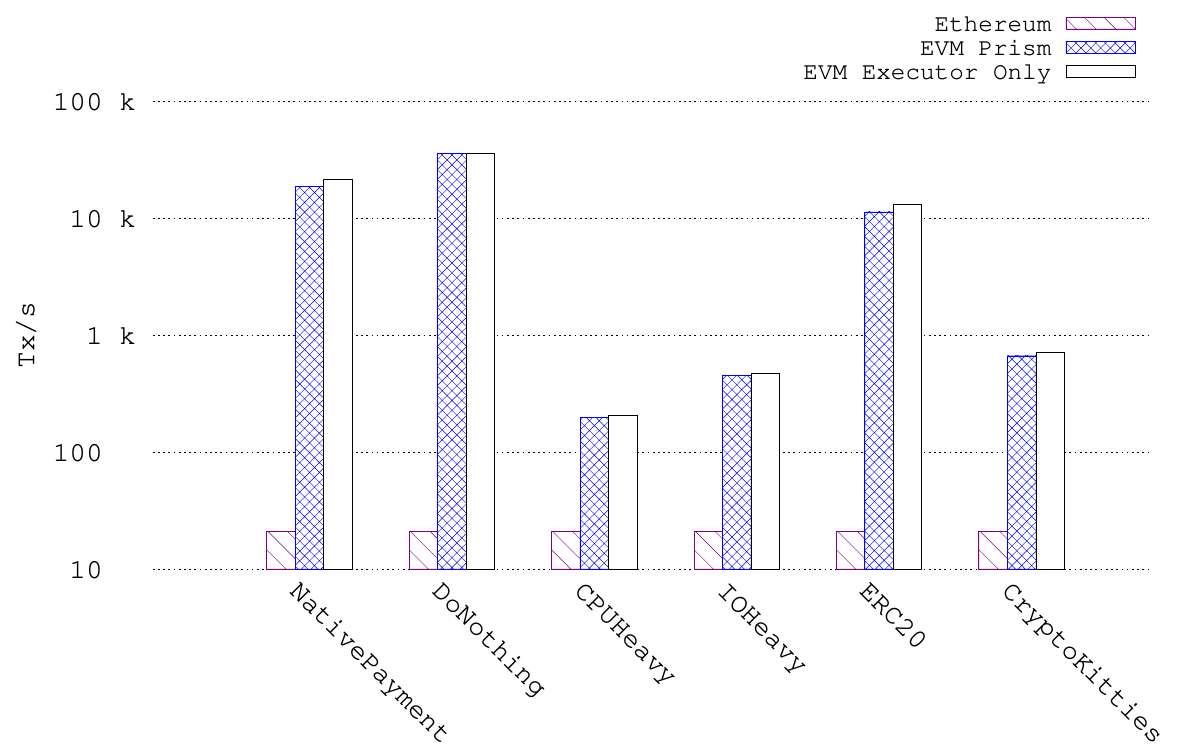}
    \caption{\evm \xspace \prism}
    \label{fig:evm_main}
    \end{subfigure}
\begin{subfigure}[b]{0.875\columnwidth}
    \centering
    \hspace{-3em}\includegraphics[width=\columnwidth]{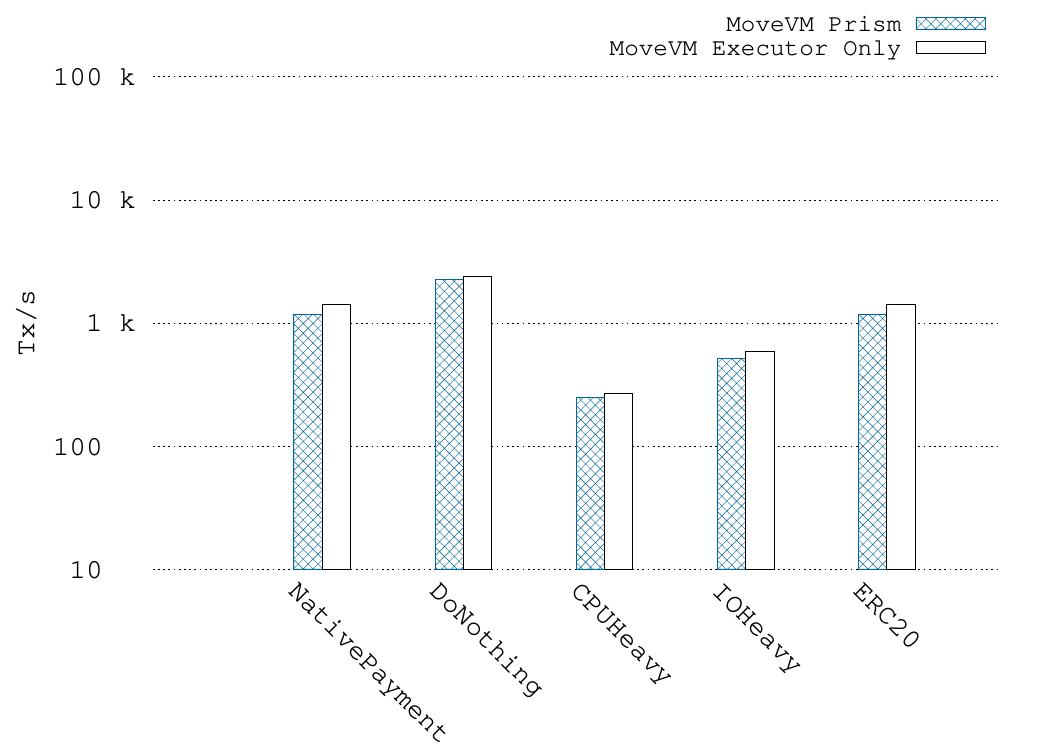}
    \caption{\movevm \xspace \prism}
    \label{fig:move_main}
    \end{subfigure}
    \caption{Throughput of \prism clients; experimented with 100 nodes on several applications: \nativepayment, \donothing, and \erc are lightweight applications whereas others are heavyweight ones. In \movevm, \nativepayment is essentially the same as \erc. The reader is cautioned against comparing performance across the two VM's, as \evm is a mature technology while \movevm is current under active development. Rather, the main point of obtaining results in two VM's is to demonstrate the flexibility of \prism. Moreover, we did not compare with the the performance of \movevm on \libra consensus because it is permissioned while \prism is permissionless. }
    \label{fig:main}
\end{figure}

\prism~\cite{prism-theory} is a recent permissionless proof-of-work (PoW) consensus protocol which naturally scales the performance of the longest chain protocol. It provably achieves throughput and latency up to computation and communication limits of the underlying physical network, while retaining the strong security guarantees of the longest chain protocol. An implementation of \prism~\cite{prism-system} scales performance significantly in a Bitcoin-like payment system, improving the throughput of \bitcoin by about $4$ orders of magnitude.  
The question remains as to whether \prism can successfully support a general smart contract platform and remove the consensus bottleneck. Indeed, not every blockchain consensus protocol is extensible to a smart contract platform (eg: Spectre \cite{sompolinsky2016spectre}) and scalably integrating consensus with smart contract platforms is nontrivial.


This paper demonstrates that \prism can support general smart contract platforms and provide a very high level of performance. We present the design and implementation of \prism that provides a flexible interface for connecting with two common smart contract virtual machines.
We report experimental results from implementation of two smart contract virtual machines, \ethereum VM (\evm) and \movevm, on top of \prism. Fig.~\ref{fig:evm_main}  shows throughput results for running several canonical smart contract applications on  \evm on \prism, while Fig.~\ref{fig:move_main} shows analogous results for \movevm on \prism. As can be seen, the throughputs are very close to that of virtual machine execution only {\em without consensus}, and much larger than the throughput using the longest chain protocol. Thus, we conclude that smart contract platforms built on \prism can perform without the consensus layer bottleneck. 


The rest of this paper is structured as follows. In \cref{sec:related} we discuss smart contract scaling approaches in different dimensions. \cref{sec:overview} gives a brief overview of \prism consensus protocol. In \cref{sec:implement}, we describe our design and implementation of \prism with \evm and \movevm. We present evaluation results of various canonical applications in \evm and \movevm, and discuss their implications in \cref{sec:eval}. Conclusion is in \cref{sec:conclusion}.

\section{Related Work}\label{sec:related}
The throughput of blockchains with smart contract platform can be increased at three different points on the blockchain stack. The first approach is to improve the execution speed of the virtual machine engine. A basic approach is to optimize the execution of individual op codes (followed in EVM  clients such as Parity \ethereum and Geth) or by  designing a new set of op codes from first principles  (followed by Libra to arrive at MoveVM~\cite{blackshear2019move}). 
A more involved approach  is to execute smart contracts in {\em parallel} similar to the modern design of databases such as MySql~\cite{mysql} and Postgres~\cite{postgres}. The first technique is to run multiple smart contracts  in parallel where smart contracts acquire locks on a data before editing to ensure no data is simultaneously edited by more than a single smart contract; this method is used in~\cite{dickerson2017adding}, with a 33\% improvement in throughput. An alternative approach uses optimistic concurrency with rollbacks; here multiple smart contracts execute in parallel (without locks) and in the case when two smart contracts running in parallel try to edit the same data, one of them is rolled backed and executed later; this approach is explored in~\cite{anjana2019efficient,saraph2019empirical,pang2019concurrency,bartoletti2019true} where 3-4x improvement in throughput is observed. Although the improvement in  throughput is significant in these methods, it exposes the blockchain to new kinds of adversarial attacks. Moreover these methods don't address metering which is a critical component to align incentives.


Even though the current VMs have low throughput, the current bottleneck in today's blockchain platform is the consensus protocol itself. Longest chain protocol and its current variants do not saturate the performance of the underlying VMs (refer Fig.~\ref{fig:main} for details). Therefore, the second approach of designing high throughput consensus protocols is a natural avenue to scale smart contract platforms. One method is to move from permissionless to permissioned consensus protocols which can support high throughput, and Facebook's Libra~\cite{libra} and IBM's Hyperledger Fabric~\cite{androulaki2018hyperledger} take this path. Libra has chosen a recent  high-performing Byzantine Fault Tolerance (BFT) consensus protocol (HotStuff~\cite{hotstuff}); Hyperledger Fabric~\cite{androulaki2018hyperledger} proposes the execute-order-validate paradigm in order to attain both performance and extensibility, where (1) participants execute transactions and collect endorsements for the executions, (2) responsible participants order these executed transactions through a consensus protocol, and (3) transactions are validated by all participants. However, these approaches sacrifice the very important characteristic of being permissionless.  In this paper we take the approach of designing and implementing a high throughput {\it permisionless} consensus protocol, \prism, which achieves high throughput. Protocols such as OHIE~\cite{yu2018ohie}, Algorand~\cite{gilad2017algorand}, Bitcoin-ng~\cite{eyal2016bitcoin} take a similar route. To the best of knowledge, there do not exist  implementations running smart contracts on top of these protocols; hence we have not been able to make a direct comparison with Prism's performance.

The third approach is Plasma and sharding.
In 2015, Poon and Buterin proposed Plasma~\cite{poon2017plasma}, along the lines of MapReduce,  an off-chain scaling solution. Many offshoots of plasma have been proposed by different communities and refer the following webpage~\cite{plasmaguide} for an overview. At a high level, Plasma is a network of secondary chains, each custom designed to serve different needs. These chains interact among each other and the main chain (on  a need basis) to resolve conflicts using fraud proofs. This approach has weaker security properties and, in particular, susceptible to the ``mass exit'' attack. To overcome some of these security vulnerabilities, 
Ethereum 2.0~\cite{ethereum2}, near~\cite{near}, polkadot~\cite{polkadot}, and Trifecta  take the sharding approach which \textit{horizontally} scales the throughput by running multiple instances of blockchains and pooling them to obtain high security. Even though this approach has better security than plasma, overall it has lower security compared to the  pure consensus protocols in the previous paragraph. 


\section{Overview of Prism}
\label{sec:overview}

\begin{figure}
\begin{center}
\includegraphics[width=\columnwidth,bb=0 0 1026 380]{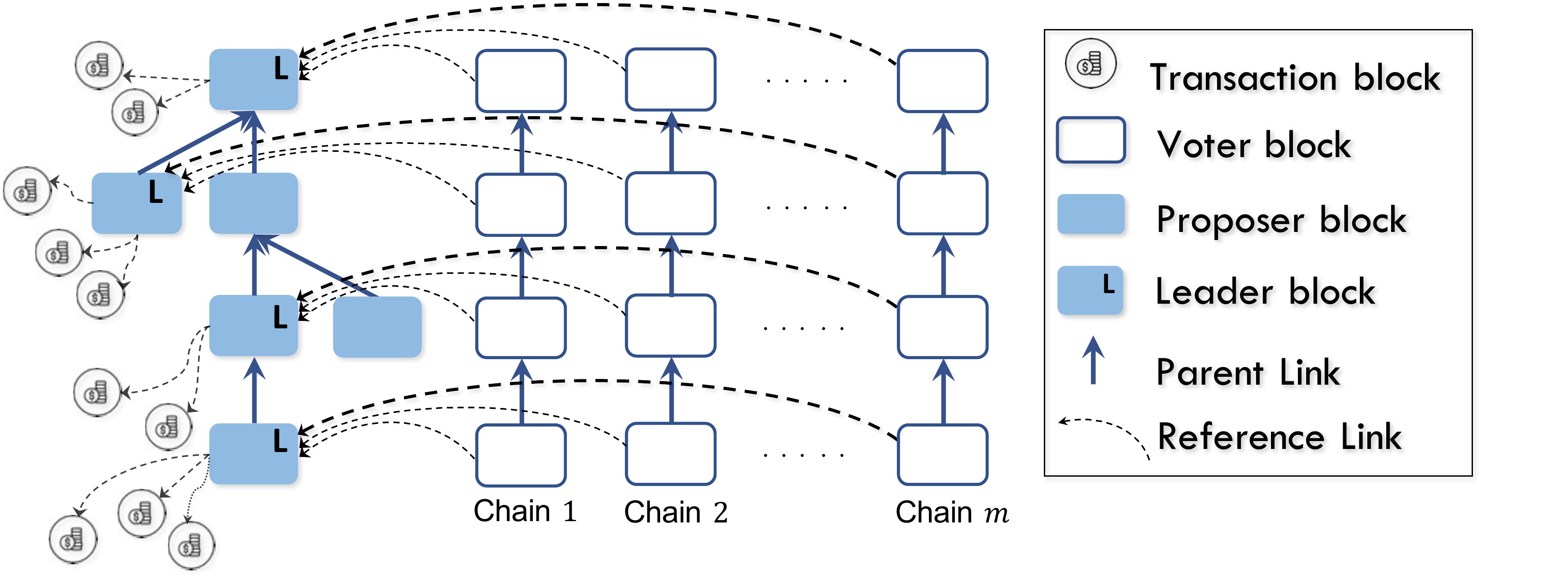}
\end{center}
\vspace{-2mm}
\caption{\small \prism: Factorizing the blocks into three types of blocks: proposer blocks, transaction blocks and voter blocks.}
\label{fig:prism}
\vspace{-5mm}
\end{figure}

The selection of a main chain in a blockchain protocol can be viewed as electing a leader block among all the blocks at each level of the blocktree. In this light, the blocks in the longest chain protocol can be viewed as serving three distinct roles: they stand for election to be leaders;  they add transactions to the main chain; they vote for ancestor blocks through parent link relationships. The latency and throughput limitations of the longest chain protocol are due to the {\em coupling} of the roles carried by the blocks. Prism removes these limitations by factorizing the blocks into three types of blocks: proposer blocks, transaction blocks, and voter blocks (Fig.~\ref{fig:prism}). Each block mined by a miner is randomly sortitioned into one of the three types of blocks, and if it is a voter block, it will be further  sortitioned into one of the voter trees. 

The proposer blocktree anchors the Prism blockchain.
Each proposer block contains a list of reference links to transaction blocks, which contains transactions, as well as a single reference to a parent proposer block.
Honest nodes mine proposer blocks on the longest chain in the proposer tree, but the longest chain does not determine the final confirmed sequence of proposer blocks, known as the  \emph{leader sequence}. 
We define the \emph{level} of a proposer block as its distance from the genesis proposer block, and the \emph{height} of the proposer tree as the maximum level that contains any proposer blocks. 
The leader sequence of proposer blocks contains one block at every level up to the height of the proposer tree, and is  determined by the \emph{voter chains}. 


There are $m$ voter chains, where $m \gg 1$ is a fixed parameter chosen by the system designer. For example, we choose $m=1000$ in our experiments.  The $i$th voter chain is comprised of voter blocks that are mined on the longest chain of the $i$th voter trees. A voter block votes for a proposer block by containing a reference link to that proposer block, with the requirements that: 1) a vote is valid only if the voter block is in the longest chain of its voter tree; 2) each voter chain votes for one and only one proposer block at each level. The leader block at each level is the one which has the highest number of votes among all the proposer blocks at the same level (tie broken by hash of the proposer blocks.) The elected leader blocks then provide a unique ordering of the transaction blocks to form the final confirmed ledger. 

By decoupling the various types of blocks, Prism can provably achieve low latency and high throughput while maintaining high security. 

\subsection{Latency}
\label{sec:prism-latency}


The votes from the voter trees secure each leader proposer block, because changing an elected leader requires reversing enough votes to give them to a different proposer block in that level. 
Each vote is in turn secured by the longest chain protocol in its voter tree. If the adversary has less than $50\%$ hash power, and the mining rate in each of the voter trees is kept small to minimize forking,  then the consistency and liveness of each voter tree guarantee the consistency and liveness of the ledger maintained by the leader proposer blocks. 
However, this would appear to require a long latency to wait for each voter block to get sufficiently deep in its chain. What is interesting is that when there are many voter chains, the same guarantee can be achieved without requiring each and every vote to have a very low reversal probability, thus drastically improving over the latency of the longest chain protocol. 

\begin{theorem}[Latency, Thm. 4.8~\cite{prism-theory}] \label{cor:latency_fast}
For an adversary with $\beta< 50\%$ of hash power, network propagation delay $D$, Prism with $m$ chains confirms  \textit{honest}\footnote{Honest transactions are ones which have no conflicting double-spent transactions broadcast in public.} transactions at reversal probability $\epsilon$ guarantee with latency upper bounded by
\begin{equation}
\label{eq:latency}
 Dc_1(\beta) +  \frac{Dc_2(\beta)}{m} \log \frac{1}{\epsilon}\;\; \text{ seconds},
\end{equation}
where $c_1(\beta)$ and $c_2(\beta) $ are $\beta$ dependent constants.
\end{theorem}

For large number of voter chains $m$, the first term dominates the above equation and therefore Prism achieves near optimal latency, i.e. proportional to the propagation delay $D$ and independent of the reversal probability.

\subsection{Throughput}
\label{sec:thruput}

To keep Prism secure, the mining rate and the size of the voter blocks have to be chosen such that each voter chain has little forking. The mining rate and the size of the proposer blocks have to be also chosen such that there is very little forking in the proposer tree. Otherwise, the adversary can propose a block at each level, breaking the liveness of the system. Hence, the throughput of Prism would be as low as the longest chain protocol if transactions were carried by the proposer blocks directly.

To decouple security from throughput, transactions are instead carried by separate transaction blocks. Each proposer block when it is mined refers to the transaction blocks that have not been referred to by previous proposer blocks. This design allows throughput to be increased by increasing the mining rate of the transaction blocks, without affecting the security of the system. The throughput is only limited by the computing or communication bandwidth limit $C$ of each node, thus potentially achieving $100\%$ utilization.  

\begin{theorem}[Throughput, Thm. 4.4~\cite{prism-theory} ]
\label{thm:throughput} For an adversary with $\beta < 50\%$ fraction of hash power and network capacity C, Prism can achieve  $(1-\beta)C$ throughput and maintain liveness in the ledger.
\end{theorem}

\section{Design and Implementation}\label{sec:implement}
We implement a \prism full-node client with VMs in around 10,000 lines of {\sf Rust} code. In this section, we describe the architecture of the client and highlight several design choices that are tailored to \prism consensus.

\subsection{Architecture}
\begin{figure}[!htb]
\centering
\includegraphics[width=0.8\columnwidth,bb=0 0 233 450]{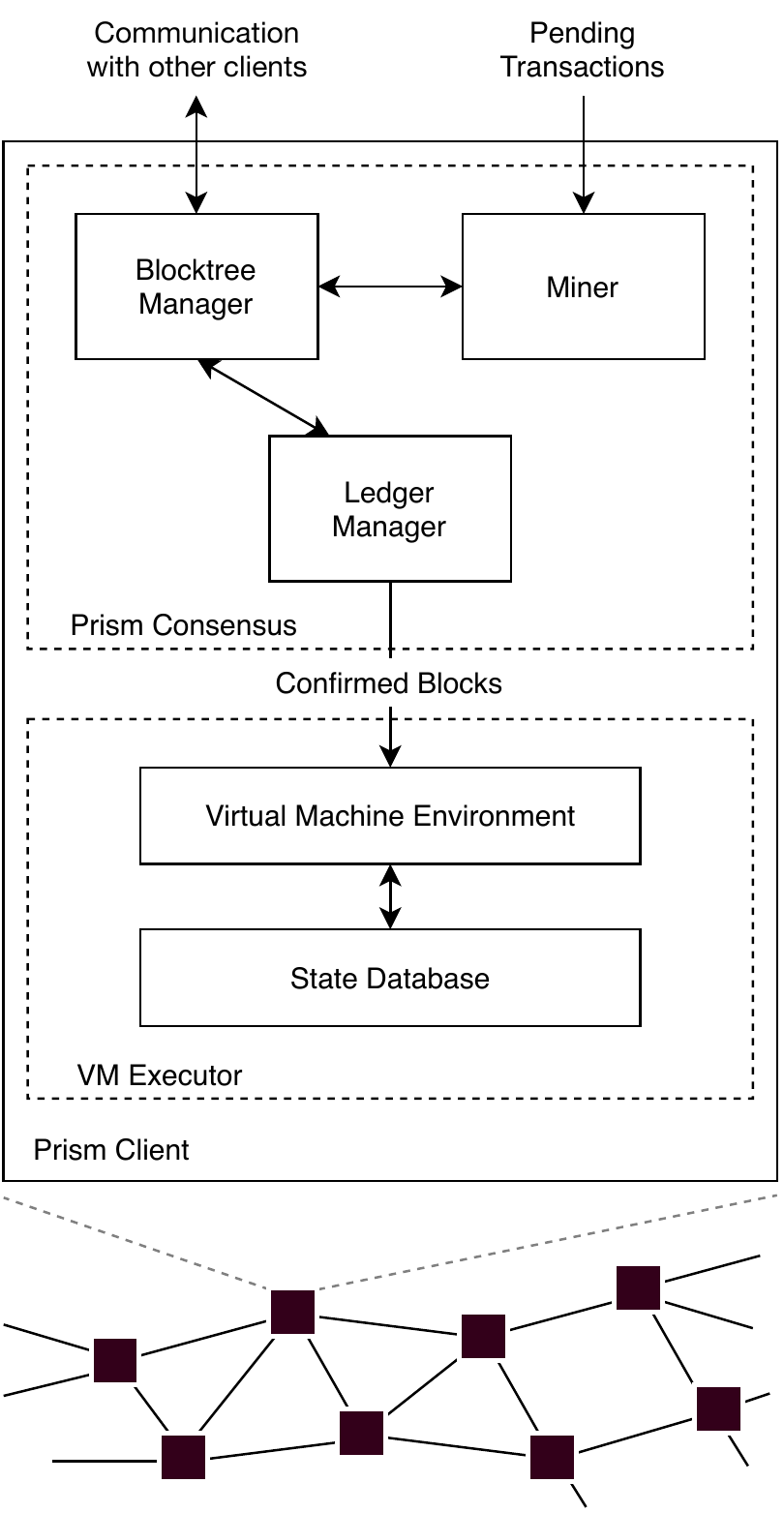}
\caption{Architecture of the \prism client. In the peer-to-peer network, each node is running a \prism full-node client.}
\label{fig:architecture}
\end{figure}

Our implementation of \prism full-node client consists of two modules, \prism Consensus module and Virtual Machine Executor (VM Executor) module. \prism Consensus module is in charge of exchanging blocks with peers, following \prism consensus to confirm blocks, and push confirmed blocks to VM Executor. VM Executor maintains the \textit{state} of the confirmed ledger, i.e., the state that results from executing transactions up to the last confirmed block. When VM Executor receives new confirmed blocks from \prism Consensus, it retrieves transactions from those blocks and updates the state accordingly. This architecture is illustrated in Fig.~\ref{fig:architecture}.

\textbf{\prism Consensus} module can be divided into the following three parts:
\begin{enumerate}
    \item Blocktree  Manager, which maintains the client's view of the blockchain, and exchanges blocks with peers;
    \item Ledger Manager, which confirms blocks by following \prism protocol, and pushes confirmed blocks to VM Executor;
    \item Miner, which contains a transaction memory pool and assembles new blocks.
\end{enumerate}

{\bf Blocktree Manager} consists of an event loop and a thread pool. The event loop keeps listening to events such as sending/receiving blocks, and assigns a thread from the thread pool to process it.
When the client receives a new block from a peer, Blocktree Manager checks its proof of work, and stores the block locally. After that, it relays the block to peers in case they have not received it. It then checks data availability, i.e., whether all the blocks referred by reference links 
in this block have been received. If not, it buffers the block and defers further processing until data availability is satisfied. After data availability is satisfied, Blocktree  Manager checks sortition and transaction signatures. 
Finally the block is inserted into \prism blocktree. 

{\bf Ledger Manager} is a busy-waiting loop that queries Blocktree Manager periodically to see whether there are new confirmed blocks, following \prism's confirmation rule. If there are, it will retrieve the blocks from local storage and push them to VM Executor via a message-passing channel. The choice of the busy-waiting loop suits the high transaction workload since the busy-waiting overhead is negligible when it takes a long time to  retrieve a large number of blocks and push them to VM Executor. Both Blocktree and Ledger Managers use RocksDB as the storage backend~\cite{rocksdb,rustrocksdb}; this choice is made due to its high performance and ease of integration.  

{\bf  Miner module} maintains  a memory pool that collects pending transactions and assembles them into new blocks. 
The Miner module does not actually try to solve the PoW hash inequality, instead  simulating the mining process by a Poisson process (of fixed growth rate, corresponding to the mining difficulty level); the Poisson processes are statistically independent across the different nodes (matching the distributed nature of PoW mining). 
When a new block is mined, it is pushed to Blocktree Manager, which will broadcast the block to peers. Transactions carried by assembled or received blocks are checked for duplication in the memory pool, with duplicates being purged.

\textbf{VM Executor} is in charge of maintaining the \textit{state database}, i.e., the persistent storage for the state of the confirmed ledger. State database stores account information such as address and balance, and manage data in a hash accumulator (Merkle Patricia tree is used in \ethereum and  sparse Merkle tree is used in \libra). 
VM Executor receives confirmed blocks from Ledger Manager, retrieves transactions from those blocks, and executes them sequentially. To execute a transaction, VM Executor first initializes a virtual machine environment, such as program counter, stack, and memory. Then it executes the instructions coded inside the transaction and/or the smart contract, during which it may interact with the state database. The execution result of a transaction will be a success or a failure, depending on whether the transaction is valid or not. Invalid transactions with failure results should be \textit{sanitized} out of the confirmed ledger and have no effect on the state. Valid transactions will update the state according to the execution result. After executing all transactions in a confirmed block, VM Executor commits the updates to the state database.

We ported the VM Executors from two open source projects, Open Ethereum~\cite{openethereum} (popularly known as Parity Ethereum) and \libra~\cite{libra}, and adapt the structure of transactions, the hash function, and the signature schemes to these projects respectively. The port only required us to add or modify less than 20 lines of code (LOC) for Open Ethereum and less than 160 LOC for \libra in their {\sf Rust} language codebases; in addition, 2 LOC were modified in Move language for \libra codebase. The two VM Executors run  single threads, with no parallel transaction execution capability. We will use the name of virtual machines, \evm and \movevm, to refer them hereafter.

\subsection{Highlights}
The key design and implementation challenge is in translating  the high throughput, low latency and high confirmation probability that \prism provides on raw block and transaction level into an application layer programming construct via the virtual machine intermediaries. On one hand, the client must process blocks and transactions at a rate much higher than most traditional blockchains. On the other hand, low latency and high confirmation probability enables confirmation of the ledger, which the implementation can benefit from. 
Here, we highlight several implementation choices that are tailored for \prism consensus and distinguish our implementation from traditional blockchains. 

\paragraph{Confirmation}
In \ethereum and other longest chain protocols, the state of the longest chain tip is used for transaction validation. However, blocks in longest chain may be switched due to honest or adversarial forking blocks. To smoothly update state when the longest chain switch happens, \ethereum's implementation  keeps a short-term journal containing actions in recent forking blocks. 
This  makes the management of state less efficient, which is a particular impediment due to the high mining rate (and high throughput) of \prism. In our design, we find it relevant to only maintain the state of the last confirmed block; this is because of two reasons: (a) \prism guarantees confirmation with overwhelmingly high probability (e.g. $1-10^{-9}$) so confirmed blocks are not likely to be deconfirmed. (b) \prism does not validate transactions before including them in blocks so it is unnecessary to maintain the state of the unconfirmed latest proposer block.  This not only makes maintenance more efficient, but also enables the integration with VM of BFT consensus such as \movevm. 

In traditional blockchains (\bitcoin and \ethereum), blocks are mined at a relatively low rate and a newly mined block is likely to change the longest chain. Hence in their implementation, they update state when they receive a new block. In \prism, blocks are mined at a high mining rate; confirming blocks and updating state upon receipt of a new block would be onerous -- we make a design choice to update the state only when blocks are confirmed and to conduct the  confirmation procedure at periodic intervals. 

\paragraph{Decoupling Transaction Validation and State Update}
In most traditional blockchains, transaction validation and state update are coupled with consensus. For example, \ethereum miners must make sure all the transactions in a block are valid, update \ethereum state accordingly, and record the result state root in that block. \prism, by design, decouples transaction validation and state update from consensus:  \prism miners do not conduct  transaction validation or  update state. Only after a block is confirmed, transactions in it are validated, and state is updated accordingly. 
In this procedure, invalid transactions are \textit{sanitized} out of the confirmed ledger. We note that invalid transactions still incur gas fees for the  senders and thus a rational user has no incentive to send invalid transactions. If the transaction sender has inadequate balance to pay the gas fee, the transaction will be treated as spam and skipped. Nevertheless this type of invalid transactions could reduce the utility of network bandwidth. To mitigate this spamming attack, miners could validate transactions (by checking sender's balance is no less than gas fee) with respect to their latest confirmed state, giving the adversary only a small window to create invalid transactions and spam the system. By this method, spam traffic is reduced by 80\% whereas the confirmation latency is only increased by 5 seconds~\cite{prism-system}. We did not implement this defense against spamming attack in this paper.

Prior work Hyperledger Fabric~\cite{androulaki2018hyperledger} also separates transaction validation and state update from consensus by the following three-step execute-order-validate paradigm: (1) nodes execute transactions and collect endorsements for the executions, (2) responsible nodes order the executed transactions through a consensus protocol, and (3) the ordered transactions are validated by all nodes and the state is updated according to valid transactions. 
\prism is similar to Hyperledger Fabric in the sense that they both separate transaction validation and state update from consensus. Notably, a common feature for \prism and Hyperledger Fabric is that the ledger could possibly contain invalid transactions at first, which would be sanitized out of the ledger later. Nevertheless, they are different in  two ways. (a) \prism orders transactions, then executes and validates them. In other words, it adopts the order-execute-validate paradigm in contrast to Hyperledger Fabric's execute-order-validate paradigm. This order-execute-validate paradigm is closely related to traditional consensus protocols, whereas the paradigm of Hyperledger Fabric  deviates far from traditional ones. (b) In Hyperledger Fabric, the execution of a transaction only occurs on a special set of nodes, and the validation requires these nodes to sign and transmit endorsements. Whereas in \prism, validation does not require such endorsements in that a transaction is validated by checking its local execution outcome; this execution and validation are replicated on every node.  
These differences imply that \prism fits well with current platforms such as \ethereum and can replace traditional consensus protocols seamlessly, while Hyperledger Fabric requires some efforts to design full-node and light-node clients in order to meet its novel requirements.


\paragraph{No Pending Transaction Exchange}
Most traditional blockchain clients exchange pending transactions in their memory pools with peers. Because the block mining rate is very low and the next block author is unpredictable, transaction exchange is necessary to ensure that pending transactions get included in the next block. This reduces network bandwidth utility since transactions are broadcast twice in the network: first as pending transactions and then as part of a block.

In \prism, pending transaction exchange can be onerous to the network bandwidth, due to the high throughput. We design our   implementation to avoid exchanging pending transactions, by noting that a pending transaction can be easily included in a {\em new} block in a very short amount of time   by any individual miner thanks to the high mining rate of \prism's transaction blocks. Transaction blocks carrying pending transactions are broadcast to peers, in the same way  blocks are broadcast in traditional blockchains. Notice that a user can still send a transaction to multiple miners for redundancy; however, miners need not exchange it. This avoids the waste of network bandwidth and contributes to the final high throughput.

\paragraph{Signature Verification in Consensus}
Transaction signature verification is a significant fraction  of total computation; this burden is only worse when the achieved throughput is higher. We design our implementation to conduct the signature verification in  {\em parallel  inside \prism Consensus} via the thread pool functionality.
This is a departure from implementations in \evm (\ethereum) and  \movevm (\libra)  which conduct signature verification inside the VM executor.  Either sequentially or in parallel, signature verification burdens the VM executor and harms the throughput. 

\section{Evaluation}\label{sec:eval}
In this section we describe our experiments and performance results of our implementation of a prototype client designed based on the guidelines highlighted in the previous section. 
We describe experiment settings and the applications that we measure (\cref{eva:01}). Then we present the throughput and confirmation latency results of \prism integrated with two virtual machines, \evm and \movevm, from which we analyze that \prism removes the consensus bottleneck (\cref{eva:02}, \cref{eva:03}). In addition, we measure how our design and implementation of \prism scales with more network participants (\cref{eva:04}).

\subsection{Experiment Setting}\label{eva:01}
We evaluate our implementation of \prism by integrating it with two smart contract virtual machines: \evmprism and \movevmprism respectively. The performance (upper  bound) baselines are provided by  \vmexecutoronly (single node, no consensus) and \textit{\prismconsensusonly} (no smart contract platform, raw transaction throughput). 
\vmexecutoronly experiment feeds transactions to VM Executor running on a single node and demonstrates the optimal throughput of the VM Executor. \prismconsensusonly experiment runs consensus with raw blocks and transactions and measures the raw data throughput. It shows the performance that the consensus is able to support. 
In addition, we also implement \ethereum's consensus protocol (essentially the longest chain protocol)  
and its performance provides a (lower bound) baseline. 

\textbf{Applications:} We evaluate a suite of canonical applications, which can be classified into three categories.

1) \textit{Basic applications:}
We evaluate two basic applications: \nativepayment and \donothing. \nativepayment transactions are payments of native tokens in those smart contract platforms.  
\donothing is a contract with a void function, and is the simplest possible contract.

2) \textit{Benchmark applications:}
To test \prism client with standard computation or storage read/write, we propose two applications: \cpuheavy and \ioheavy. \cpuheavy runs a worst case of quick sort for an integer array of length 255. \ioheavy does key-value pair write 255 times followed by key-value pair read 255 times for both forward and backward order (thus total 510 times). The value type is bytes32 in \evm and bytearray in \movevm, which are both 256-bit data type.

3) \textit{Realistic applications:}
As a counterpoint to the above applications, we  evaluate here the performance with respect to two real world applications: \erc and \cryptokitties. ERC20 is an \ethereum token standard~\cite{erc20}, and we implement it by using the reference implementation in~\cite{openzeppelin}. CryptoKitties is a game that allows users to breed virtual pets. The genes of offspring are determined by a function named \textit{mixGenes} that mixes the genes of its parents~\cite{cryptokitties}. We adopt \textit{mixGenes} function in our experiments, and feed random parent genes to it. This function is significantly computational heavy compared to basic applications.

Applications for \evm are developed in Solidity programming language. 
We use the official Solidity compiler v0.6.3 to compile all smart contracts to bytecode except for \cryptokitties, which we follow the version v0.4.18 in the contract. We set the compiler to Constantinople version and enable the default optimization. When creating a smart contract in \evm, an account address is created and bytecode is stored under the address. Applications for \movevm are developed in Move IR. The smart contracts are first published as modules under the sender's address and then are called via scripts. We use Move IR compiler to compile the modules and scripts to bytecode. We have basic applications and benchmark applications and they have the same functionality as corresponding applications for \evm. Native tokens in \movevm have essentially the same function as ERC20 tokens in \evm, hence \erc experiment for \movevm is unnecessary. 
As the Move language is in rapid development and not yet mature at the moment of our experiments, it is not straightforward to implement \cryptokitties in \movevm.

Table~\ref{tab:applications} presents the statistics of applications. Transaction sizes differ because we pass different input parameters to these applications. Number of instruction and gas are indicators of the complexity in terms of both computation and storage read/write. \movevm does not provide the statistics for number of instruction.

\begin{table}[htb]
    \centering
    \caption{\evm and \movevm application statistics.}
    \label{tab:applications}
    \begin{tabular}{ccccccc}
    \toprule
        & \makecell{Native\\Payment} & \makecell{Do\\Nothing}  & \makecell{CPU\\Heavy} & \makecell{IO\\Heavy} & ERC20 & \makecell{Crypto-\\Kitties} \\
    \midrule
        \makecell{\evm\\Tx Size\\(Bytes)} & 533 & 536  & 567 & 567& 601 & 631\\ \hline
        \makecell{\evm\\Gas} & 21000 & 21394 & 334390 & 435244  & 26602 & 140000 \footnotemark\\ \hline
        \makecell{Num of\\Instruction} & 0 & 32 & 88417 & 25364 & 309 & 25000 \footnotemark[\value{footnote}]\\
    \midrule
        \makecell{\movevm\\Tx Size\\(Bytes)} & 424 & 329 &365 & 366 & N/A & N/A \\ 
        \hline
        \makecell{\movevm\\Gas} & 43076 & 629 & 2275420 & 2956846 & N/A & N/A \\
    \bottomrule
    \end{tabular}
\end{table}
\footnotetext{Since we pass random inputs to \cryptokitties, the number is also random and we present an approximation in the table.}

To generate the workloads for our evaluations, we implement a transaction generator that periodically generates transactions and push them into the mempool,  generating different transaction types for different applications. We cap the generation rate according to the throughput of \vmexecutoronly experiment, in order not to exhaust the virtual machine.

We acquire data from the first 100 million transactions on \ethereum to derive a distribution on the number of transactions sent and received by an account. We sample our transactions using this distribution to mimic the usage of \ethereum in our experiments. In our experiments we use 10,000 accounts in total for both sender and receiver. The transaction generator of each node is initialized with 10,000 key pairs; one key pair for each account. In order to mimic the usage of \ethereum for \nativepayment and \erc, each node randomly and independently draws a sender and a receiver address from the aforementioned distribution. Other applications like \donothing, \cryptokitties, \cpuheavy, and \ioheavy have a fixed receiver (\evm) or no receiver (\movevm) and hence we only sample the sender address.

\textbf{Experiment environment}. We perform our experiments on Amazon EC2's 100 c5d.4xlarge instances. Each instance has 16 CPU cores, 32 GB memory, and NVMe SSD storage. Each instance hosts one \prism client and they are connected to form a random 4-regular topology; the diameter of the network is 6. To emulate a realistic peer-to-peer network, we introduce a propagation delay of 120 ms on each link to match the typical delay in \ethereum's network~\cite{gencer2018decentralization}, and a rate limiter of 300 Mbps for both ingress and egress traffic, except for \prismconsensusonly experiment where the rate limiter is 600 Mbps in order to show the performance upper bound that the consensus can reach.

\textbf{Parameters}.
For \evmprism and \movevmprism, we choose a high adversarial hash power capability of  $\beta=0.4$ and a very low deconfirmation probability $\epsilon=2\times 10^{-9}$. We use $m=1000$ voter chains and cap the size of transaction blocks to be 200 tx/block. Given the testbed with 120 ms peer-to-peer delay, we tune the mining rate of \prism's proposer and voter blocks to be 0.08 block/s, at which the empirical forking rate~\footnote{Forking rate is calculated by $1-\frac{\text{\# blocks in longest chain}}{\text{\# blocks}}$.} is less than 0.11 in all experiments, and thus it ensures the security of \prism. We tune the mining rate of transaction blocks differently for different applications to match  the throughput of \vmexecutoronly experiment: 
In \evmprism, \nativepayment 108; \donothing 180; \erc 70; \cryptokitties 3.78; \cpuheavy 1.08; \ioheavy 2.34 block/s. In \movevmprism, \nativepayment 12.6; \donothing 7.2; \cpuheavy 1.44; \ioheavy 3.06 block/s.

For the \prismconsensusonly experiment, we increase the size of transaction blocks to 400 tx/block and the mining rate to 200 block/s in order to show the performance upper bound that the consensus can reach. As for the \textit{\ethereum} experiment, we use a mining rate of 0.1 block/s and a block size of 200 tx/block, which resemble the live \ethereum parameters.

\begin{figure}[hbt]
    \centering
    \includegraphics[width=\columnwidth]{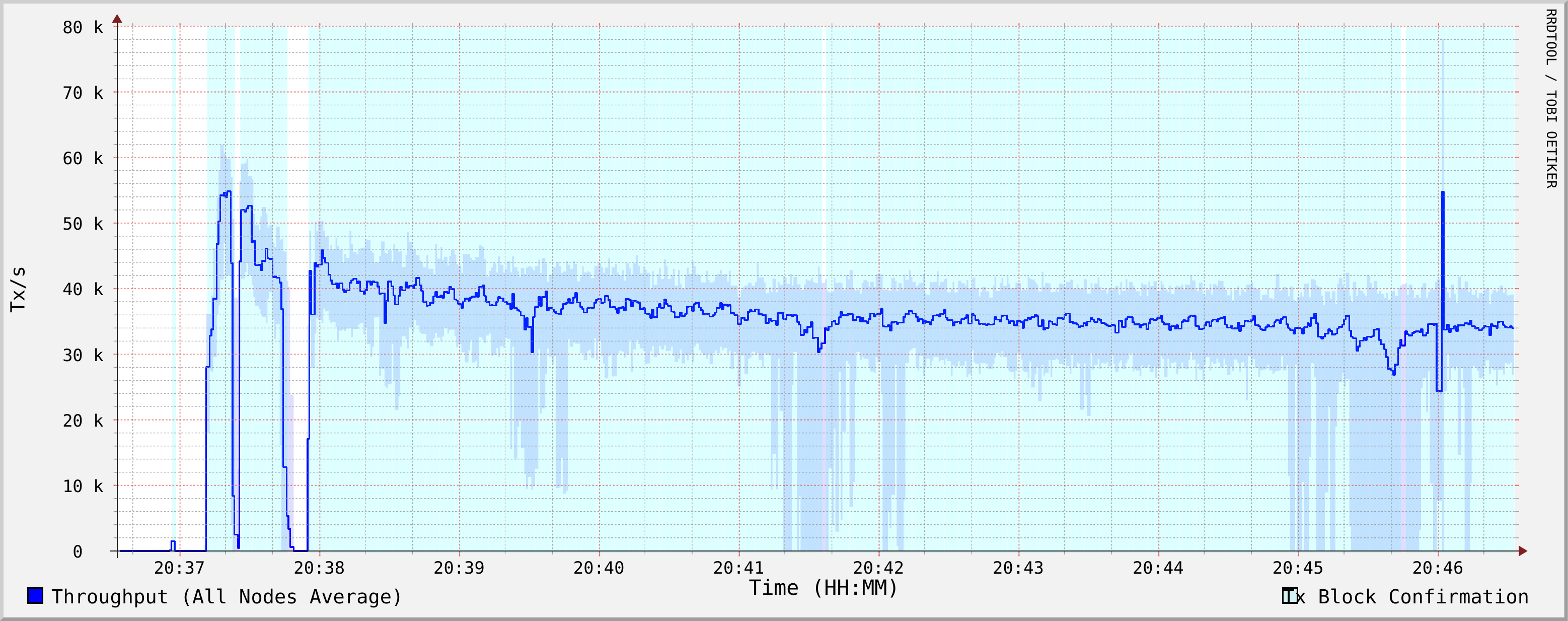}
    \caption{Time series plot of the throughput for \donothing application in \evmprism for 10 minutes. Around the first 30 seconds there are very few processed transactions, since the clients are just started and have not extended the ledger significantly. After the ledger starts to be extended significantly, the throughput soon increases and becomes stable. This phenomenon occurs in all \prism experiments.}
    \label{fig:time_series}
\end{figure}

All experiments are run for at least 10 minutes. As we  see in the time series plot of the throughput (Fig.~\ref{fig:time_series}), in the first several seconds, the nodes don't process any transaction because they just started mining blocks and there are not enough blocks to extend the confirmed ledger. This phenomenon only happens at the beginning and does not affect the performance afterwards. Hence, the final  throughput calculation involves the average performance over the last 9 minutes of the experiment.

\subsection{Throughput and Latency of \evm}\label{eva:02}
In this experiment, we measure the transaction throughput  and confirmation latency of various applications in \evmprism and analyze the difference in throughput for different applications. 
We also compare the throughput with \evmexecutoronly experiment, the optimal throughput of \evm on a single node. If the former is able to reach the latter, then the throughput of our \prism client is very close to the optimal throughput of the virtual machine and we can conclude that  \prism removes the consensus bottleneck for smart contracts.
Finally  we compare \evmprism with \prismconsensusonly to study whether \prism is able to support even higher throughput without the limitation of the virtual machine. This experiment would also indicate whether \evmprism's performance can be further improved   if the underlying virtual machine becomes faster.

\begin{table}[!th]
    \centering
    \caption{Throughput in terms of tx/s on \evm applications.}
    \label{tab:01}
    \begin{tabular}{ccccccc}
    \toprule
        & \makecell{Native\\Payment} & \makecell{Do\\Nothing} & \makecell{CPU\\Heavy} & \makecell{IO\\Heavy} & ERC20 & \makecell{Crypto-\\Kitties} \\
    \midrule
        \makecell{\evm\\Executor\\ Only}& 21535 & 35723 & 207 & 467 & 13095 & 710 \\ \hline
        \makecell{\evm\\\prism} & 18660 & 35329 & 197 & 447 & 11210 & 661 \\\hline
        \makecell{\prism\\Consensus\\Only~\cref{consensus_only}} & 98022 & 97473 &92144&92144 &86931&82798\\\hline
        \ethereum & \multicolumn{6}{c}{21}\\
    \bottomrule
    \end{tabular}
\end{table}
\stepcounter{footnote}
\footnotetext{\label{consensus_only}In \prismconsensusonly, the consensus throughput is 418 Mbps and is converted to tx/s based on transaction size.} 

\textbf{Throughput:} As shown in Table~\ref{tab:01}, for \evmprism, the throughput of two basic applications is able to reach 18K and 35K tx/s respectively. For \erc, \evmprism gets 11K tx/s. The throughput of these three applications shows that we have a good chance to get above ten thousand tx/s for those applications that do not involve heavy computation or storage read/write. The reason that \donothing is almost twice as fast as \nativepayment is that for \donothing, the VM Executor module updates the account information of a random sender per transaction and a fixed receiver contract account, whereas for \nativepayment it updates a random sender and a random receiver account information. As a result, the VM Executor needs to maintain the state database and hash accumulator for half account information updates in \donothing as that in \nativepayment.

For the  \cryptokitties application,  \evmprism achieves 661 tx/s due to its computational heavy nature. Similar things happen for \cpuheavy and \ioheavy, which get 197 and 447 tx/s respectively. According to the statistics in Table~\ref{tab:applications}, these applications require more than 25K instructions in the virtual machine, which explains their low throughput. However, the low throughput in both \evmprism and \evmexecutoronly also indicates that \evm has a large opportunity to improve the efficiency of execution. We write exactly the same \cpuheavy application in {\sf Java} and run in JVM, and we get a throughput over 90K tx/s. Considering the large gap between 197 and 90K, we believe that \evm has the potential to execute instructions more efficiently. We don't compare \ioheavy or \cryptokitties since they are not as straightforward to implement  as a standalone program in {\sf Java}.

\textbf{Is \prism consensus the bottleneck?}
For all \evm applications, \evmprism reaches 85\% of \evmexecutoronly throughput. This high percentage indicates that \evmprism is able to reach the optimal \evm throughput very closely. As for \prismconsensusonly, we can see the high throughput over 80K tx/s for all applications. This high number illustrates the ability of supporting a high throughput without the limitation of the virtual machine. It also shows that if the virtual machine becomes faster in the future, \prism is able to support its performance as well. Hence, \prism consensus is not the throughput bottleneck;  the  virtual machine itself is the bottleneck.

Compared to 21 tx/s in \textit{\ethereum} experiment, which adopts Nakamoto's longest chain consensus, it is clear that the current \ethereum is limited by consensus. 

\textbf{Latency:}
The end to end  latency of a transaction consists of two parts: confirmation latency and execution latency. Confirmation latency is the time between a transaction is generated and the corresponding block is confirmed. This latency is decided by \prism's confirmation rule and has a proved bound~\cite{prism-theory}. Execution latency is the time that a transaction waits in a queue to be executed and the time of execution. As long as we cap the transaction generation rate below the optimal virtual machine throughput in experiments, the time in the queue is negligible. Also the execution time is less than ten milliseconds since all applications have over one hundred tx/s throughput. Hence, execution latency is negligible compared to confirmation latency.

\prism's confirmation rule guarantees a confirmation latency regardless of its throughput.
In all \prism experiments including \evmprism, \movevmprism, and \prismconsensusonly, the confirmation latency is no more than 130 seconds. Notice that this latency is achieved with adversarial ratio $\beta=0.4$ and reversal probability $\epsilon=2\times 10^{-9}$. To provide the same latency under the same condition in \ethereum, it needs to wait for ($k=267$)-deep~\cite{bitcoin} and it translates to 2670 seconds if a block is mined in 10 seconds on average.

\textbf{Resource utility:}
In a \prism client, the \prismconsensus module uses multiple threads to process messages from/to peers efficiently. The VM Executor module, on the contrary, runs in a single thread. In addition, RocksDB uses a few threads in the background. In total, a \prism client should only use no more than 50 threads. In our experiments, the live usage of CPU never exceeds 50\% per core on average (notice that one instance has 16 CPU cores). Though, there are possible optimization to do in the future. For example, give a high priority to the VM Executor thread to prevent competing CPUs with the \prismconsensus module. 

By profiling the CPU usage of a client in \evmprism \xspace \donothing experiment, we find that transaction signature verification takes up to 39.2\% of total CPU time (excluding mining), a relatively high percentage. When the experiment is running at a high throughput, the requirement of a large amount of signature verification is a major bottleneck; this emphasizes the importance of removing signature verification from the VM Executor module. In our design, we have moved signature verification into the \prismconsensus module, thus freeing the VM Executor from this heavy burden.

The VM Executor of \evm is implemented efficiently with abundant number of in-memory cache. However, it levies a heavy memory burden on the VM Executor; 
our port of \evm does not include the whole client-level cache management, as a result, the VM Executor does not free memory efficiently, and the memory usage increases along with the workload. This is one possible future optimization for our port of \evm.

Table~\ref{tab:breakdown} provides a breakdown statistics for three \prism block types in \evmprism \xspace \donothing experiment. We can see that transaction blocks take up to 71.2\% of total generated block data, other two blocks only 28.8\%. This indicates that the majority of utilized bandwidth contributes to the high throughput (transaction blocks), whereas \prism overhead takes up only a small fraction (proposer and voter blocks). For other \evmprism (and \movevmprism) experiments, this breakdown statistics will remain similar except for transaction blocks. The higher the throughput, the higher the transaction block data and percentage. Thus, we do not analyze the breakdown statistics for other experiments.
\begin{table}[thb]
    \centering
    \caption{Statistics for three block types in \evmprism \xspace \donothing experiment in 10 minute duration.}
    \label{tab:breakdown}
    \begin{tabular}{cccc}
    \toprule
         & \# Mined Block & Block Data & Data Percentage \\
    \midrule
        Proposer Block & 44  & 4.0 MB& 4.0\% \\
        Voter Block &  47166  & 25.2 MB&24.8\% \\
        Transaction Block &  107514  & 72.2 MB &71.2\% \\
    \bottomrule
    \end{tabular}
\end{table}

\subsection{Throughput and Latency of \movevm}\label{eva:03}
In this experiment, we measure the transaction throughput and confirmation latency for \movevmprism. We observe similar bottleneck and latency between this experiment and \evm experiment, whereas there are also discrepancies in terms of throughput.

\textbf{Throughput:}
As shown in Table~\ref{tab:02}, the throughput of two basic applications is only 1.1K and 2.2K tx/s respectively; this is an order of magnitude smaller than that of \evmprism. In private communication~\cite{private-communication-sam}, core \libra developers have indicated to us that improving the performance of \movevm is  work under progress -- when this improvement transpires, our \prism client can fully utilize that as well.  
Benchmark applications get 249 and 512 tx/s and are higher than those of \evmprism, indicating that \movevm is more efficient at executing instructions. The \cpuheavy throughput number, however, is still far below that of JVM (over 90K tx/s), so we believe \movevm has the  potential to execute instructions  more efficiently as well.

\begin{table}[!th]
    \centering
    \caption{Throughput in terms of tx/s on \movevm applications.}
    \label{tab:02}
    \begin{tabular}{ccccc}
    \toprule
        & \makecell{Native\\Payment} & \makecell{Do\\Nothing} & \makecell{CPU\\Heavy} & \makecell{IO\\Heavy} \\
    \midrule
        MoveVM Executor Only & 1441 & 2501 & 269 & 585\\
        MoveVM Prism & 1172 & 2243 & 249 & 512 \\
        Prism Consensus Only~\cref{consensus_only} &123222&158802&143140&142749\\
    \bottomrule
    \end{tabular}
\end{table}


\textbf{Is \prism consensus the bottleneck?}
For all \movevm applications, \movevmprism reaches 81\% of \movevmexecutoronly throughput. This phenomenon is similar to \evm and indicates that \movevmprism is able to reach the optimal \movevm throughput. As for \prismconsensusonly, we can see the high throughput over 120K tx/s as well. Similar to the case of \evm, we conclude that \prism removes the consensus bottleneck for \movevm, and the virtual machine itself is the bottleneck.

\textbf{Latency:}
\prism guarantees a confirmation latency regardless of the throughput, and we do observe that in all \prism experiments including \movevmprism, the average confirmation latency is no more than 130 seconds.

\textbf{Resource utility:}
\movevmprism maintains a good memory usage, which is kept under 3.2 GB in all experiments. The live usage of CPU never exceeds 32\% per core on average; compared to \evmprism experiment, this CPU utility reduction is due to smaller throughput and more efficient signature verification. \movevm adopts Ed25519 signature~\cite{bernstein2012high} which is faster than ECDSA~\cite{johnson2001elliptic} adopted by \evm. 

\subsection{Scalability}\label{eva:04}
In this experiment, we evaluate \prism's ability to scale with more network participants. We use a larger number, 300, EC2 instances and use the same propagation delay and rate limiter. We use a random 5-regular topology for 300 nodes, keeping diameter the same with that of 100 nodes. We also keep the same \prism parameter, including the overall mining rate, thus the individual mining rate is modified. By our design, only the \prismconsensus module is related to scaling with more network participants, since only it communicates with peers. In addition, \prismconsensus module's performance is not affected by which application it is running. Hence, it suffices to experiment with one VM and application to demonstrate \prism's scalability and we use \evmprism and \nativepayment in the experiment.

The experiment for 300 nodes also runs for 10 minutes. However, it is hard to collect the fine-grained metrics for such a high number of nodes. So we calculate the overall metrics at the end of the experiment (all 10 minutes), in contrast to previous calculation (last 9 minutes).

Table~\ref{tab:scale} compares the performance between 100 and 300 nodes. The throughput and latency are very similar; the difference is due to the randomness of the experiments. The forking rate  0.113 in 300 nodes is a little larger than that in 100 nodes, and is mainly due to more hops and higher delay to propagate blocks throughout the peer-to-peer network, as we can see that the average path length is higher in the 300-node topology. This forking rate 0.113 is small enough to ensure the security of \prism consensus as well. In addition, the block propagation delay, as well as the forking rate, can be reduced by increasing the degree (the number of peers per node) of the peer-to-peer network; Geth~\cite{geth} and Parity Ethereum~\cite{openethereum} client have a default maximum degree of 50, which can sustain a low forking rate for peer-to-peer networks with a larger number of nodes. 

\begin{table}[htb]
    \centering
    \caption{Performance of \evmprism \xspace \nativepayment, with different network topologies.}
    \label{tab:scale}
    \begin{tabular}{ccccccc}
    \toprule
        \#Node & Degree &\makecell{Average\\Path\\Length} & Diameter & Throughput & \makecell{Confirmation\\Latency\\(s)}& Forking \\
    \midrule
        100 & 4 & 3.55 & 6 & 17268 & 96 & 0.102 \\
        300 & 5 & 3.84 & 6 & 17417 & 80 & 0.113 \\
    \bottomrule
    \end{tabular}
\end{table}

Resource utility on each node is also similar. In the 300-node experiment, the live usage of CPU never exceeds 50\% per core on average. The heavy memory burden of the VM Executor module is also similar to that in 100-node experiment.

We conclude that \prism is able to scale to a large number of network participants, as long as the underlying peer-to-peer network provides a topology with reasonable block propagation delay. We can achieve similar throughput, latency, and security in those cases.

\section{Conclusion}\label{sec:conclusion}

Blockchain research thus far has progressed in a compartmentalized manner: algorithms and protocols (many focused on consensus)  are designed and studied separately from the upper layer wrappers (virtual machine, application programming) they will interact with. This is in contrast with Nakamoto's \bitcoin design that was envisioned and designed as a complete system.  This layering philosophy works well when the consensus layer is the bottleneck and much work can be expended to improve the performance (indeed, this is the case with many blockchains, including \ethereum). \prism is a recent consensus algorithm, closely inspired by Nakamoto's longest chain protocol,  with {\em theoretically  optimal}   throughput and latency.  In this paper we explore how Prism fits with two smart contract virtual machines, \evm and \movevm, by implementing \prism underneath these virtual machines. We demonstrate that \prism seamlessly merges with both these VMs:  our implementation approaches the {\em optimal} virtual machine throughput for a large variety of applications. This result means that \prism not only removes the consensus bottleneck of bare metal throughput and latency but also when interacting with  two popular smart contract platforms.  Further improvement of the smart contract performance would have to come from new designs of virtual machines and compilers and architectures capable of parallel execution of smart contracts. The early  research in this area \cite{saraph2019empirical,dickerson2019adding} now takes on added urgency. 






\section*{Acknowledgement}
We  thank Lei Yang for his core role in the development of the \prism codebase, which we have  integrated with \evm and \movevm in this paper. We thank Sam Blackshear and  Andrew Miller for helpful discussions.   This research is supported in part by NSF under grants CCF-1705007, NeTS-1718270 and Army Research Office under grant W911NF-17-S-0002. 
\bibliographystyle{IEEEtran}
\bibliography{references}

\begin{thebibliography}{10}
\providecommand{\url}[1]{#1}
\csname url@samestyle\endcsname
\providecommand{\newblock}{\relax}
\providecommand{\bibinfo}[2]{#2}
\providecommand{\BIBentrySTDinterwordspacing}{\spaceskip=0pt\relax}
\providecommand{\BIBentryALTinterwordstretchfactor}{4}
\providecommand{\BIBentryALTinterwordspacing}{\spaceskip=\fontdimen2\font plus
\BIBentryALTinterwordstretchfactor\fontdimen3\font minus
  \fontdimen4\font\relax}
\providecommand{\BIBforeignlanguage}[2]{{%
\expandafter\ifx\csname l@#1\endcsname\relax
\typeout{** WARNING: IEEEtran.bst: No hyphenation pattern has been}%
\typeout{** loaded for the language `#1'. Using the pattern for}%
\typeout{** the default language instead.}%
\else
\language=\csname l@#1\endcsname
\fi
#2}}
\providecommand{\BIBdecl}{\relax}
\BIBdecl

\bibitem{prism-theory}
V.~Bagaria, S.~Kannan, D.~Tse, G.~Fanti, and P.~Viswanath, ``Prism:
  Deconstructing the blockchain to approach physical limits,'' in
  \emph{Proceedings of the 2019 ACM SIGSAC Conference on Computer and
  Communications Security}, ser. CCS ’19, 2019, p. 585–602.

\bibitem{bitcoin}
S.~Nakamoto, ``Bitcoin: A peer-to-peer electronic cash system,'' 2008.

\bibitem{optimisticrollup}
\BIBentryALTinterwordspacing
\emph{Optimism}. [Online]. Available: \url{https://optimism.io/}
\BIBentrySTDinterwordspacing

\bibitem{zkrollup}
\BIBentryALTinterwordspacing
\emph{ZK-Rollups}. [Online]. Available:
  \url{https://docs.ethhub.io/ethereum-roadmap/layer-2-scaling/zk-rollups/}
\BIBentrySTDinterwordspacing

\bibitem{kalodner2018arbitrum}
H.~Kalodner, S.~Goldfeder, X.~Chen, S.~M. Weinberg, and E.~W. Felten,
  ``Arbitrum: Scalable, private smart contracts,'' in \emph{27th $\{$USENIX$\}$
  Security Symposium ($\{$USENIX$\}$ Security 18)}, 2018, pp. 1353--1370.

\bibitem{prism-system}
L.~Yang, V.~Bagaria, G.~Wang, M.~Alizadeh, G.~Fanti, D.~Tse, , and
  P.~Viswanath, ``Prism: Scaling bitcoin by 10,000 $\times$,''
  \emph{arXiv:1909.11261}, 2019.

\bibitem{sompolinsky2016spectre}
Y.~Sompolinsky, Y.~Lewenberg, and A.~Zohar, ``Spectre: A fast and scalable
  cryptocurrency protocol.'' \emph{IACR Cryptology ePrint Archive}, vol. 2016,
  p. 1159, 2016.

\bibitem{blackshear2019move}
S.~Blackshear, E.~Cheng, D.~L. Dill, V.~Gao, B.~Maurer, T.~Nowacki, A.~Pott,
  S.~Qadeer, D.~R. Rain, S.~Sezer \emph{et~al.}, ``Move: A language with
  programmable resources,'' 2019.

\bibitem{mysql}
\BIBentryALTinterwordspacing
\emph{My Sql}. [Online]. Available: \url{https://www.mysql.com/}
\BIBentrySTDinterwordspacing

\bibitem{postgres}
\BIBentryALTinterwordspacing
\emph{PostGres}. [Online]. Available: \url{https://www.postgresql.org/}
\BIBentrySTDinterwordspacing

\bibitem{dickerson2017adding}
T.~Dickerson, P.~Gazzillo, M.~Herlihy, and E.~Koskinen, ``Adding concurrency to
  smart contracts,'' in \emph{Proceedings of the ACM Symposium on Principles of
  Distributed Computing}, 2017, pp. 303--312.

\bibitem{anjana2019efficient}
P.~S. Anjana, S.~Kumari, S.~Peri, S.~Rathor, and A.~Somani, ``An efficient
  framework for optimistic concurrent execution of smart contracts,'' in
  \emph{2019 27th Euromicro International Conference on Parallel, Distributed
  and Network-Based Processing (PDP)}.\hskip 1em plus 0.5em minus 0.4em\relax
  IEEE, 2019, pp. 83--92.

\bibitem{saraph2019empirical}
V.~Saraph and M.~Herlihy, ``An empirical study of speculative concurrency in
  ethereum smart contracts,'' \emph{arXiv preprint arXiv:1901.01376}, 2019.

\bibitem{pang2019concurrency}
S.~Pang, X.~Qi, Z.~Zhang, C.~Jin, and A.~Zhou, ``Concurrency protocol aiming at
  high performance of execution and replay for smart contracts,'' \emph{arXiv
  preprint arXiv:1905.07169}, 2019.

\bibitem{bartoletti2019true}
M.~Bartoletti, L.~Galletta, and M.~Murgia, ``A true concurrent model of smart
  contracts executions,'' \emph{arXiv preprint arXiv:1905.04366}, 2019.

\bibitem{libra}
\BIBentryALTinterwordspacing
\emph{Libra}, accessed February 16, 2020. [Online]. Available:
  \url{https://github.com/libra/libra}
\BIBentrySTDinterwordspacing

\bibitem{androulaki2018hyperledger}
E.~Androulaki, A.~Barger, V.~Bortnikov, C.~Cachin, K.~Christidis, A.~De~Caro,
  D.~Enyeart, C.~Ferris, G.~Laventman, Y.~Manevich \emph{et~al.}, ``Hyperledger
  fabric: a distributed operating system for permissioned blockchains,'' in
  \emph{Proceedings of the Thirteenth EuroSys Conference}, 2018, pp. 1--15.

\bibitem{hotstuff}
M.~Yin, D.~Malkhi, M.~K. Reiter, G.~G. Gueta, and I.~Abraham, ``Hotstuff: Bft
  consensus with linearity and responsiveness,'' in \emph{Proceedings of the
  2019 ACM Symposium on Principles of Distributed Computing}, ser. PODC ’19,
  2019, p. 347–356.

\bibitem{yu2018ohie}
H.~Yu, I.~Nikolic, R.~Hou, and P.~Saxena, ``Ohie: Blockchain scaling made
  simple,'' in \emph{2020 IEEE Symposium on Security and Privacy (SP)}.\hskip
  1em plus 0.5em minus 0.4em\relax IEEE Computer Society, may 2020, pp.
  112--127.

\bibitem{gilad2017algorand}
Y.~Gilad, R.~Hemo, S.~Micali, G.~Vlachos, and N.~Zeldovich, ``Algorand: Scaling
  byzantine agreements for cryptocurrencies,'' in \emph{Proceedings of the 26th
  Symposium on Operating Systems Principles}, 2017, pp. 51--68.

\bibitem{eyal2016bitcoin}
I.~Eyal, A.~E. Gencer, E.~G. Sirer, and R.~Van~Renesse, ``Bitcoin-ng: A
  scalable blockchain protocol,'' in \emph{13th $\{$USENIX$\}$ symposium on
  networked systems design and implementation ($\{$NSDI$\}$ 16)}, 2016, pp.
  45--59.

\bibitem{poon2017plasma}
J.~Poon and V.~Buterin, ``Plasma: Scalable autonomous smart contracts,''
  \emph{White paper}, pp. 1--47, 2017.

\bibitem{plasmaguide}
\BIBentryALTinterwordspacing
\emph{Plasma}. [Online]. Available:
  \url{https://ethresear.ch/t/plasma-world-map-the-hitchhiker-s-guide-to-the-plasma/4333}
\BIBentrySTDinterwordspacing

\bibitem{ethereum2}
\BIBentryALTinterwordspacing
\emph{Ethereum 2.0 Specifications}. [Online]. Available:
  \url{https://github.com/ethereum/eth2.0-specs}
\BIBentrySTDinterwordspacing

\bibitem{near}
\BIBentryALTinterwordspacing
\emph{NEAR Protocol | A sharded, developer-friendly, proof-of-stake public
  blockchain}. [Online]. Available: \url{https://nearprotocol.com/}
\BIBentrySTDinterwordspacing

\bibitem{polkadot}
\BIBentryALTinterwordspacing
\emph{Polkadot: Decentralized Web 3.0 Blockchain Interoperability Platform}.
  [Online]. Available: \url{https://polkadot.network/}
\BIBentrySTDinterwordspacing

\bibitem{rocksdb}
\BIBentryALTinterwordspacing
\emph{RocksDB}. [Online]. Available: \url{https://rocksdb.org/}
\BIBentrySTDinterwordspacing

\bibitem{rustrocksdb}
\BIBentryALTinterwordspacing
\emph{rust-rocksdb}, accessed February 19, 2020. [Online]. Available:
  \url{https://github.com/rust-rocksdb/rust-rocksdb}
\BIBentrySTDinterwordspacing

\bibitem{openethereum}
\BIBentryALTinterwordspacing
\emph{OpenEthereum}, accessed January 30, 2020. [Online]. Available:
  \url{https://github.com/openethereum/openethereum}
\BIBentrySTDinterwordspacing

\bibitem{erc20}
\BIBentryALTinterwordspacing
\emph{ERC-20 Token Standard}. [Online]. Available:
  \url{https://github.com/ethereum/EIPs/blob/master/EIPS/eip-20.md}
\BIBentrySTDinterwordspacing

\bibitem{openzeppelin}
\BIBentryALTinterwordspacing
\emph{OpenZeppelin Contracts}. [Online]. Available:
  \url{https://github.com/OpenZeppelin/openzeppelin-contracts/blob/master/contracts/token/ERC20/ERC20.sol}
\BIBentrySTDinterwordspacing

\bibitem{cryptokitties}
\BIBentryALTinterwordspacing
\emph{CryptoKitties GeneScience}. [Online]. Available:
  \url{https://etherscan.io/address/0xf97e0a5b616dffc913e72455fde9ea8bbe946a2b#code}
\BIBentrySTDinterwordspacing

\bibitem{gencer2018decentralization}
A.~E. Gencer, S.~Basu, I.~Eyal, R.~Van~Renesse, and E.~G. Sirer,
  ``Decentralization in bitcoin and ethereum networks,'' in \emph{International
  Conference on Financial Cryptography and Data Security}.\hskip 1em plus 0.5em
  minus 0.4em\relax Springer, 2018, pp. 439--457.

\bibitem{private-communication-sam}
S.~Blackshear, {Private Communication}, 2020.

\bibitem{bernstein2012high}
D.~J. Bernstein, N.~Duif, T.~Lange, P.~Schwabe, and B.-Y. Yang, ``High-speed
  high-security signatures,'' \emph{Journal of cryptographic engineering},
  vol.~2, no.~2, pp. 77--89, 2012.

\bibitem{johnson2001elliptic}
D.~Johnson, A.~Menezes, and S.~Vanstone, ``The elliptic curve digital signature
  algorithm (ecdsa),'' \emph{International journal of information security},
  vol.~1, no.~1, pp. 36--63, 2001.

\bibitem{geth}
\BIBentryALTinterwordspacing
\emph{Go Ethereum}, accessed May 31, 2020. [Online]. Available:
  \url{https://github.com/ethereum/go-ethereum}
\BIBentrySTDinterwordspacing

\bibitem{dickerson2019adding}
T.~Dickerson, P.~Gazzillo, M.~Herlihy, and E.~Koskinen, ``Adding concurrency to
  smart contracts,'' \emph{Distributed Computing}, pp. 1--17, 2019.

\end{thebibliography}
\end{document}